\documentclass[aps,prstab,twocolumn,10pt,groupedaddress]{revtex4-2}
\usepackage{tikz}
\usepackage{graphicx}
\usepackage{url}
\usepackage{fancyvrb}
\usepackage{color}
\usepackage{tikz}
\usepackage{hyperref}
\usepackage{float}
\usepackage{ulem}
\usepackage{todonotes}

\usetikzlibrary{shapes,arrows,positioning,shapes.geometric}
\usepackage{charter}
\usepackage{siunitx}
\usepackage{amssymb, amsmath, bm}
\usepackage{graphicx}
\usepackage{color}
\usepackage{afterpage}

\usepackage{struktex}

\usepackage{xspace}

\newcommand{\htp}{\texorpdfstring{\ensuremath{\mathrm{H}_2^+}\xspace}{H2+}}

\newcommand{\BE}[0]{\begin{equation}}
\newcommand{\EE}[0]{\end{equation}}
\newcommand{\BEA}[0]{\begin{eqnarray}}
\newcommand{\EEA}[0]{\end{eqnarray}}

\mathchardef\mhyphen="2D

\newcommand{\figref}[1]{FIG.~\ref{#1}}
\newcommand{\tabref}[1]{TAB.~\ref{#1}}

\begin{document}


\title{Input Beam Matching and Beam Dynamics Design Optimizations of the IsoDAR RFQ using Statistical and Machine Learning Techniques}


\author{Daniel Koser}
\email[]{dkoser@mit.edu}
\affiliation{Massachusetts Institute of Technology, 77 Massachusetts Ave, 
             Cambridge, MA 02139, USA}
\author{Andreas Adelmann}
\affiliation{Paul Scherrer Institut, Forschungsstrasse 111, Villigen, Switzerland}
\author{Janet Conrad}
\affiliation{Massachusetts Institute of Technology, 77 Massachusetts Ave, 
             Cambridge, MA 02139, USA}
\author{Matthias Frey}
\affiliation{Paul Scherrer Institut, Forschungsstrasse 111, Villigen, Switzerland}
\affiliation{current affiliation: Mathematical Institute, University of St Andrews, St Andrews KY16 9SS, UK}
\author{Loyd Waites}
\author{Daniel Winklehner}
\affiliation{Massachusetts Institute of Technology, 77 Massachusetts Ave, 
             Cambridge, MA 02139, USA}


\date{\today}

\begin{abstract}
We present a novel machine learning-based approach to generate fast-executing 
virtual radiofrequency quadrupole (RFQ) particle accelerators using surrogate 
modelling. These could potentially be used as on-line feedback tools during 
beam commissioning and operation, and to optimize the RFQ beam dynamics design prior to construction.
Since surrogate models execute orders of magnitude faster than corresponding 
physics beam dynamics simulations using standard tools like PARMTEQM and RFQGen, 
the computational complexity of the multi-objective optimization problem reduces
significantly. Ultimately, this presents a computationally inexpensive and time 
efficient method to perform sensitivity studies and an optimization of the crucial 
RFQ beam output parameters like transmission and emittances. Two different methods of surrogate model creation (polynomial chaos expansion and neural networks) are discussed and the achieved model accuracy is evaluated for different study cases with gradually increasing complexity, ranging from a simple FODO cell example to the full RFQ optimization. We find that variations of the beam input Twiss parameters can be reproduced
well. The prediction of the beam with respect to hardware changes, e.g.~of the electrode modulation, are
challenging on the other hand. We discuss possible reasons.
\end{abstract}


\maketitle


\section{Introduction}
\label{sec:intro}
Artificial Intelligence (AI) and Machine Learning (ML), using statistical methods and 
Neural Networks (NNs), are quickly becoming a staple of modern computational physics.
Their highly successful application in computer vision \cite{NIPS2015_14bfa6bb, 7780459}
and the establishment of many software packages that are widely available and 
standardized (e.g., TensorFlow \cite{tensorflow2015-whitepaper} and Keras \cite{chollet2015keras}) has led to attempts to use ML in almost all fields of science.
Particle accelerator physics is no exception, although ML is not as
well-established here as in other fields.
A few examples of ML in accelerator physics are given in the following.
Arguably, the best-established use of ML is image analysis using 
convolutional neural networks (CNNs). CNNs can be used in beam diagnostics
for the analysis of the output of emittance scanners, optical fibers, 
residual gas monitors, and reconstruction of beam pulse structure~\cite{ren:cnn1}.
The SwissFEL was tuned using Bayesian optimization 
\cite{kirschner:bayesian1, kirschner:bayesian2}.
Bayesian optimization, using Gaussian Process models was also used 
for the Linac Coherent Light Source (LCLS)~\cite{duris:bayesian1}.
Another very promising technique, that is also the subject of this paper,
is surrogate modelling.
We describe the method in detail later. In short, a fast-executing model of a 
complex system can be produced by training a NN or using Polynomial Chaos Expansion
(PCE) on a set of high-fidelity simulations. This fast-executing Surrogate Model (SM) can
then be used in an optimization scheme or for on-line feedback during run-time.
Some examples of successful use of surrogate models in particle accelerator 
optimization are 
REFS. \cite{adelmann:surrogate1, van_der_veken:ml1, edelen:ml1, edelen:ml2}, 
which have demonstrated speedups of one to several orders 
of magnitude compared to conventional techniques.

To our knowledge, AI/ML has not yet been applied to the design of radiofrequency 
quadrupole (RFQ) linear accelerators. Here we report our recent results
using surrogate modeling to create virtual RFQ models that can be used in 
several ways:
\begin{itemize}
\item Uncertainty Quantification (UQ) \cite{adelmann:surrogate1} 
      of the RFQ with respect to input beam
variations or RFQ settings during run-time. 
\item Prediction of output beam parameters from a given set of input beam parameters.
      The SM becomes a virtual accelerator, ideal as tuning and 
      commissioning aid.
\item Design and optimization of the RFQ hardware. Based on the success as a 
      virtual accelerator, we also tested the SM technique as a hardware optimization
      tool.
\end{itemize}
The findings in this paper are fully transferable to other RFQs.

\subsection{Particle Physics Motivation for this Work}
\begin{table}[b!]
  \centering
  \caption{Basic parameters of the IsoDAR-RFQ, corresponding to the baseline beam
  dynamics design and the preliminary RF/mechanical design.\label{RFQ-Parameters}}
  \vspace{1mm}
  \begin{tabular}{lc}
					RF frequency [MHz] & 32.8 \\
					design ion & ${\text{H}_2}^+$ \\
					design beam current [mA] & 6.5 \\
					duty cycle & cw \\
					input/output energy [keV] & 15\,/\,$\sim$70 \\
					inter-vane voltage [kV] & 20.1 \\
					beam transmission [\%] & 97.3 \\
					trans.~input emittance [$\pi$\,mm\,mrad] & 0.30 \\
					trans.~output emittance [$\pi$\,mm\,mrad] & 0.34 \\
					long.~output emittance [$\pi$\,keV\,deg] & 40.2 \\ 
					tank diameter [cm] & 28 \\
					electrode length [cm] & 136.5 \\
					RF power [kW] & $\sim$3.6 \\
					shunt impedance [k$\Omega$m] & 154 \\
  \end{tabular}
\end{table}
The motivation for this work lies in the IsoDAR project 
\cite{bungau:isodar1, abs:isodar_cdr1, winklehner:nima}, a proposed 
search for exotic neutrinos. These are hypothesized cousins to the three known
standard model neutrinos and could explain anomalies seen in the neutrino
oscillation experiments of the past three decades~\cite{diaz_where_2019}.

To reach discovery-level sensitivity ($>5\sigma$) in five years of
running, IsoDAR rquires a 10~mA cw proton beam at 60~MeV on a neutrino
production target. This accelerator (described in Refs. 
\cite{calanna:dic, winklehner:nima, winklehner:njp}) accelerates \htp
instead of protons and uses a novel RFQ direct injection method
\cite{winklehner:rfq1, winklehner:nima}, in which
the beam is aggressively pre-bunched in an RFQ that is embedded axially
in the cyclotron yoke and brought very close to the cyclotron
median plane. Because of the high beam current, necessarily small diameter
(as little yoke iron as possible must be removed), and the difficult 
matching of the RFQ output to the cyclotron acceptance, we have 
initiated this study to accurately predict the sensitivity of the RFQ,
the output beam parameters, and to optimize the RFQ design beyond
the current baseline. In \tabref{RFQ-Parameters}, we list the most 
important parameters of the IsoDAR RFQ, some of which will be used
as design variables (DVARS) and objectives (OBJ) in the reported study.

\subsection{The Structure of this Paper}
We have structured this manuscript into Methodology, Results, and Discussion.
In each section, we describe our work separately for the two applications
of the SM: 1. As Tuning and Commissioning Tool; 
2. As Design and Optimization Tool. These are the natural applications
due to the immense speedup of SMs compared to high-fidelity 
Particle-In-Cell (PIC) simulations.
We also present results for a very simple system -~the FODO cell~-
as a benchmark and to elucidate the basic principles and challenges.
In the Results, we show that the SM performs excellently as a tuning tool, but
issues arise when we vary the hardware parameters of the RFQ.
In the Discussion, we explain why we think the surrogate model
under-performs when the longitudinal beam dynamics is affected through
hardware (design parameter) changes.

\subsubsection{The Surrogate Model as Tuning and Commissioning Tool}
The first application we present is using the SM as an on-line feedback tool
during the commissioning and running of the RFQ direct injection prototype. 
We envision the SM to provide valuable assistance for the operator to allow quick or automated adjustment of the RFQ and beamline settings with respect to the input beam properties. To this end, in the
final application, we will train the SM using simulated input values like the
signal of beam position monitors (BPMs), the beam current 
(from an AC Current Transformer \cite{bergoz:acct}), and beam size (from a wire probe)
before the RFQ and predict the signals from similar devices after the RFQ.
To test the idea in this manuscript, we use the Twiss parameters of the beam as
input.

\subsubsection{The Surrogate Model as Design and Optimization Tool}
Finding an optimized beam dynamics design often requires a very large number of 
simulation iterations. This makes the design procedure of RFQs time consuming, 
especially when completely new solutions to meet the required beam output quality 
need to be explored. 
This is sometimes even the case for comparatively fast executing beam dynamics codes 
like PARMTEQM \cite{crandall:parmteq1} or RFQGen \cite{rfqgenref}, but is definitely a 
problem when time consuming PIC simulations are used as the basis for optimization.
Similar to demonstrated successes with cyclotrons and electron accelerators 
\cite{edelen:ml1, PhysRevAccelBeams.22.054602}, we are investigating the use of SMs
to perform multiobjective optimization for the RFQ modulation cell parameters, in order to
find the minimal beam output emittances (transverse and longitudinal) and maximum transmission.

\section{Methodology}
\label{sec:methods}

\subsection{Surrogate Modeling}
Surrogate models are cheap alternatives to reduce the computational complexity of multiobjective optimizations as already shown in the
context of particle accelerators in \cite{PhysRevAccelBeams.23.044601}. We chose neural networks and polynomial chaos expansions to replace the high-fidelity
RFQ model codes. These methods are explained in the following subsections. More
detailed introductions can be found in the listed references and the references contained therein.

\subsubsection{Polynomial Chaos Expansion}
The principle of the polynomial chaos expansion (PCE) relies on the orthogonality of the multivariate polynomials $\Psi_{i}$.
The high-fidelity model $m(\bm{x})$ with input vector $\bm{x}\in\mathbb{R}^{d}$ and $d\ge 1$ is approximated by
\begin{equation*}
    m(\bm{x}) \approx \hat{m}(\bm{x}) = \sum_{i=1}^{P}c_{i}\Psi_{i}(\bm{\xi})
                                      = \sum_{i=1}^{P}c_{i}\prod_{j=1}^{d}\psi_{j}(\xi_j)
\end{equation*}
where
\begin{equation*}
    P = \frac{(p+d)!}{p!d!}
\end{equation*}
is the total number of monomials determined by the expansion truncation order $p$ and the dimensionality
of the system $d$. The vector $\bm{\xi} = (\xi_{1},\dots,\xi_{d})$ represents the input vector that is mapped onto the support of the
univariate polynomials $\psi_{j}$. The type of the univariate polynomials of the $j$th dimension depends on the distribution of the
corresponding input dimension. For example, uniformly distributed dimensions are approximated by Legendre polynomials and normally
distributed dimensions by Hermite polynomials.

There are multiple methods to obtain the expansion coefficients $c_{i}$ with different requirements on the number of training points
$N$. Commonly used methods are orthogonal projection, regression and Bayesian.
In the case of the projection method, the number of training points grows exponentially with the dimension, i.e.,
\begin{equation*}
    N = (p+1)^{d}.
\end{equation*}
Regression and Bayesian approaches have no strict requirements, but according to \cite{SUDRET2008964} an optimal number of samples is given by
\begin{equation*}
    N = (d - 1) P.
\end{equation*}
A benefit of PCE based surrogate models is the evaluation of Sobol' indices \cite{Sobol01}, a measure of global sensitivity of the output on
the input. The first-order Sobol' index, also known as main sensitivity, quantifies the effect of a single input dimension. The total
effect of an input dimension, that also includes all correlations with other dimensions, is denoted as total sensitivity.

We also refer the interested reader to the following literature \cite{adelmann-2019-1} (and the references therein). Many PCE literature references can also be found in the bibliography of \cite{FREY2021107577}.

\subsubsection{Artificial Neural Networks}
 The term ``Artificial Neural Network'' (ANN) refers to a broad class of methods within Machine Learning (ML) that share the common property of consisting of many interconnected processing units that are used to transform data. The first of such a hierarchy of layers, consists of an affine linear function $T:\mathbb{R}^n\rightarrow \mathbb{R}^m$, defined as $T(x):=Wx+b$, where $W=(a_{ij})\in \mathbb{R}^{m\times n}$, $x\in \mathbb{R}^{n}$, $b\in \mathbb{R}^m$, and $n,m \in \mathbb{N}$. $W$ and $b$ are commonly referred to as the weights and biases of the ANN. The second is an activation function $\sigma:\mathbb{R}\rightarrow \mathbb{R}$, which is typically nonlinear. Many variants of $\sigma$ exist, in this work we use the rectified linear unit  $\sigma(x)=\max(0,x)$. 
 The activation function is applied in an element-wise manner, hence a vector activation function $\sigma: R^n\rightarrow R^n$ can be defined.
 Now we are able to define a continuous function $f(x)$ by a composition of
linear transforms $T^{i}$ and activation functions $\sigma$, i.e.\
 \begin{equation}\label{eqn:eg3layernet}
 f(x)=T^{k}\circ\sigma\circ T^{k-1}\circ
 \cdot\cdot\cdot~\sigma\circ T^{1}\circ\sigma\circ T^{0}(x),
 \end{equation}
 with $T^{i}(x)=W_ix+b_i$. $W_i$ are initially undetermined matrices and $b_i$ initially undetermined vectors and $\sigma(\cdot)$ is the element-wise activation function. The values of $W_i$ and $b_i$ are randomly initialized and adjusted during ``training'' using an optimization algorithm to maximize some performance metric. 
 

 Such an ANN is called a $(k+1)$-layer ANN, which has $k$ hidden layers. Denoting all the undetermined coefficients (e.g., $W_i$ and $b_i$) in \eqref{eqn:eg3layernet} as $\theta\in\Theta$, where $\theta$ is a high dimensional vector and $\Theta$ is the span of $\theta$, the ANN representation of a continuous function can now be viewed as
 \begin{align*}\label{eqn:solution_DNN}
 f=f(x;\theta).
 \end{align*}
 Let $\mathbb{F}=\{ f(\cdot,\theta)|\theta\in\Theta\}$ denote the set of all expressible functions by the ANN parameterized by $\theta\in\Theta$, then $\mathbb{F}$ provides an efficient way to represent unknown continuous functions.

 Approximation properties of neural network can be found in~\cite{Cybenko1989,HornikSW89}, where the authors studied approximation properties for the function classes given by a feed-forward neural network with a single hidden layer. In later works, authors studied the error estimates for such neural networks in terms of hyper-parameters such as number of neurons, layers of the network, and activation functions, a review can be found in~\cite{ellacott_1994} and~\cite{pinkus_1999}.

 \subsection{Data Generation for Surrogate Modelling}

\begin{figure}[h!]
   \centering
   \includegraphics[width=\columnwidth]{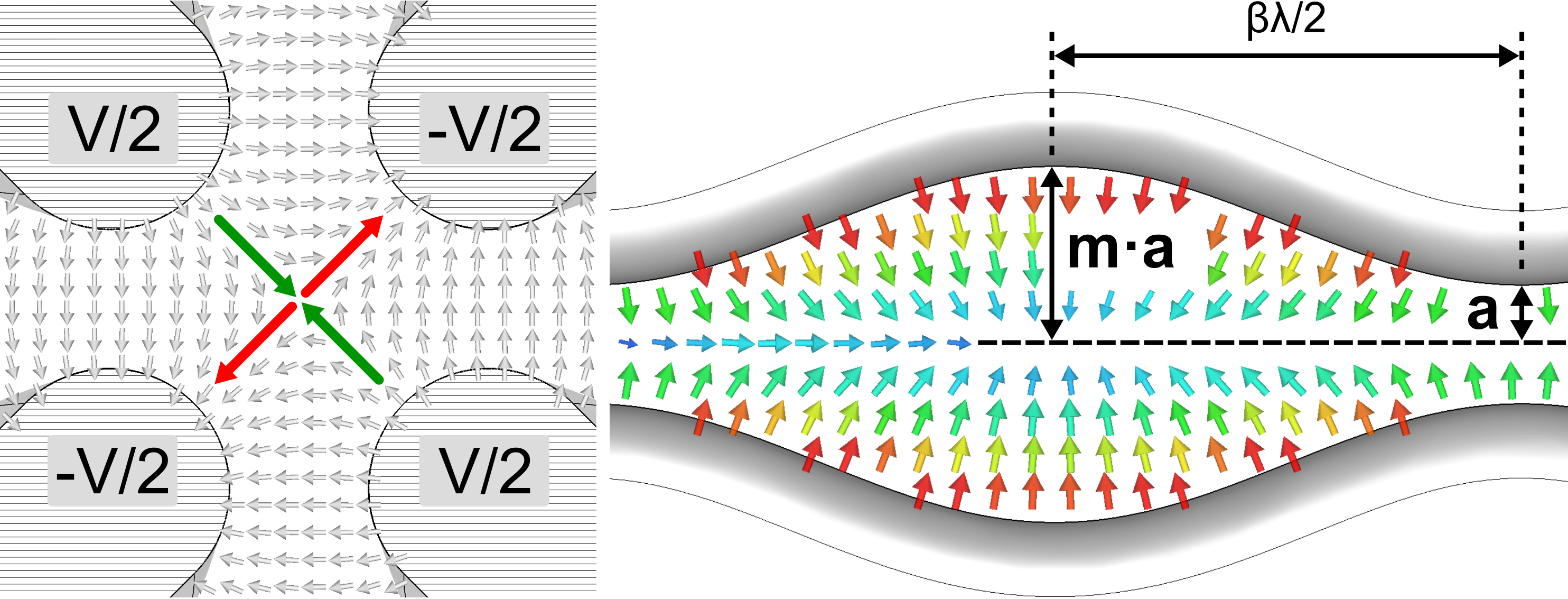}
   \caption{Transverse electric quadrupole field around the beam axis of an RFQ (\textit{left}) with focusing/defocusing plane (green/red) and electrode cell modulation (\textit{right}), resulting in a longitudinal field component.} 
   \label{fig:RFQ_schematic}
\end{figure}

The beam dynamics properties of an RFQ with a number of $n$ modulation cells are fully described by the parameter sets $\mathbf{B} = (B_1,..,B_n)$, $\mathbf{m} = (m_1,..,m_n)$ and $\boldmath{\phi_\text{s}}$~$= (\phi_{\text{s},1},..,\phi_{\text{s},n})$, quantifying the basic functions of an RFQ as descriptively explained in the following:

\begin{itemize}
    \item The transversely defocusing effect of the space charge force has a $1/\gamma^2$-dependency ($\gamma$ being the Lorentz factor) and hence at low beam velocities efficient and velocity-independent transverse focusing is required. As shown in \figref{fig:RFQ_schematic}, the alternating electric quadrupole field between the RFQ electrodes leads to a focusing force along one of the transverse axes while defocusing occurs in the perpendicular direction, effectively constituting an alternating gradient focusing channel. The transverse focusing strength in an RFQ cell $n$ is commonly characterized by the parameter $B_n$ \cite{crandall:rfq}.
    \item By adding a sinusoidal modulation to the electrode shape, a longitudinal field component is generated which can be used to adiabatically bunch the DC input beam. This is a highly delicate procedure due to the high sensitivity of space-charge dominated beams to perturbations of the beam particle density. The consecutive modulation cells form a $\pi$-mode accelerator structure with a cell length of $\ell_c = \beta_c \lambda_{\text{RF}}/2$. The extent of electrode modulation (corresponding to the magnitude of the longitudinal field component) of a cell $n$ is parameterized by the modulation factor $m_n$.
    \item The synchronous phase $\phi_{\text{s},n}$, which is set by the cell lengths, determines the ratio of longitudinal bunching to acceleration and hence the overall phase space stability. By increasing $\phi_{\text{s},n}$ along the RFQ, beam acceleration is gradually introduced.
\end{itemize}

Ultimately, the beam output properties depend on the RFQ hardware specifications as well as on the given input beam parameters, which for a DC input beam are specified by the transverse emittances, the Twiss parameters and the beam current.

 \subsubsection {Simulated Data for a Fixed RFQ Design}
 To investigate the capability of surrogate models to reproduce the RFQ beam output properties as a function of only the adjustable beam input parameters (in our case the Twiss parameters $\alpha$ and $\beta$), we used a fixed preliminary optimized RFQ design, through which we simulated the beam using the PARMTEQM code. A sample data set was obtained from the output of a number of PARMTEQM simulations with randomized values for the input Twiss parameters (corresponding to the design variables of the underlying optimization problem) within a predefined range of $\alpha = [1, 4]$ and $\beta = [7, 25] \, \text{(cm/mrad)}$. The transverse and longitudinal output emittances as well as the transmission (constituting the optimization objectives) were evaluated directly at the end of the RFQ electrodes.   
 
\subsubsection{Simulations of Full RFQ Design}
    \begin{figure*}[t!]
        \centering
        \includegraphics[width=0.95\textwidth]{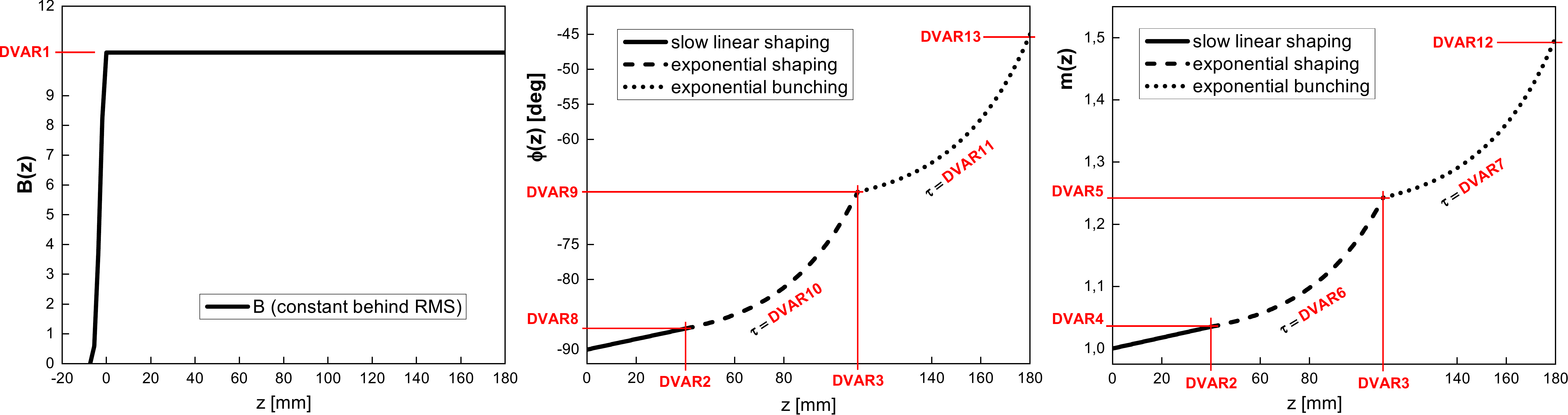}
        \caption{Parametrization functions for the RFQ cell properties 
                 ($B(z)$, $\phi(z)$, $m(z)$) specified by design variables
                 (DVARs).\label{fig:RFQparametrization}}
\end{figure*}
 To study the applicability of surrogate models for optimizing the RFQ design itself, we introduced a parameterization of the functions for transverse focusing $B(z)$, synchronous phase $\phi(z)$ and electrode modulation $m(z)$ according to \figref{fig:RFQparametrization}. This reduces the size of the RFQ design parameter space, corresponding to the number of design variables, from $3n+1$ ($B_n$, $\phi_{\text{s},n}$, $m_n$ for each cell $n$, $+1$ because the number of cells is a design variable itself) to a total number of 14.\\ 
 The parameterization functions were chosen so that the crucial properties of the underlying baseline design remain variable for optimization; e.g. the constant value of $B(z)$ behind the RMS (DVAR1), the lenghts of the linear and exponential shaping and bunching sections (DVAR2 and DVAR3) as well as the rate and smoothness of shaping and bunching (DVAR5--13). The length of the RFQ is determined by DVAR14, being the cutoff energy after which PARMTEQM ends the electrode (always with a full RFQ cell).\\
 We generated a sample data set from beam dynamics simulations using PARMTEQM for a number of random RFQ design variations (randomized DVAR values within a predefined range) with a fixed input beam (constant Twiss parameters). 
 
 \subsection{ML Training and Use of RFQ Surrogate Models}
\begin{figure*}[t!]
   \centering
    \includegraphics[width=0.8\textwidth]{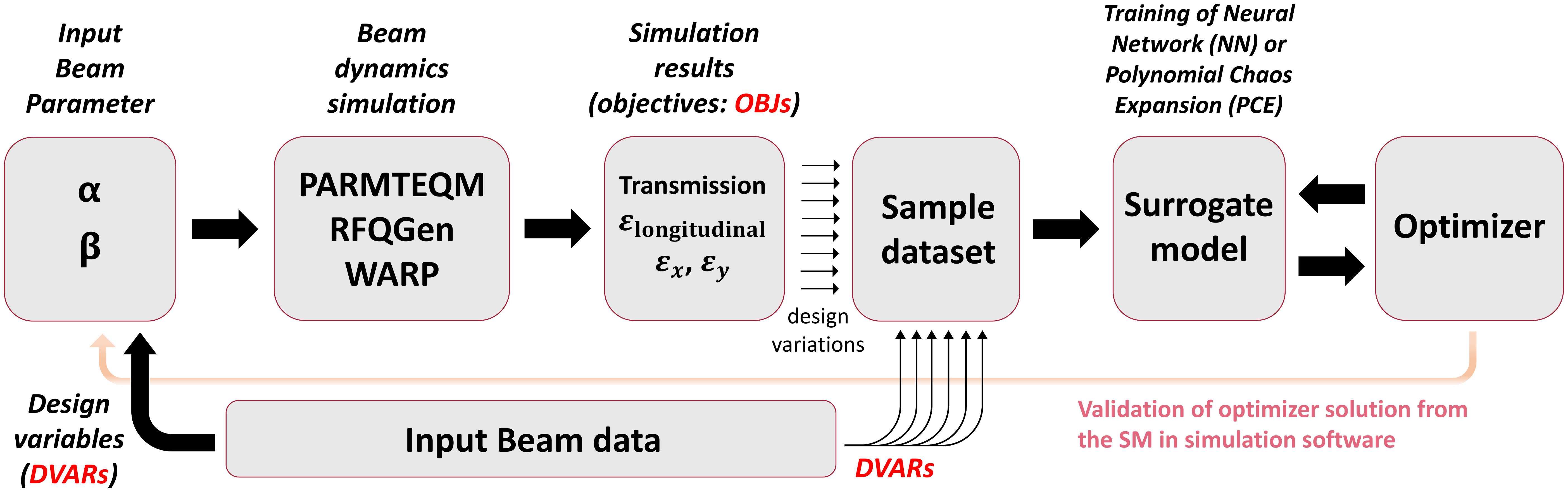}
    \caption{General machine learning optimization scheme for RFQ beam
             dynamics.\label{fig:ML_schematic_inputbeam}}
\end{figure*}

 As being best practice for the training of ML models, we randomly split sample data sets into 70\% training and 30\% test data. The training data is then used to train either a PCE or DNN surrogate model. Once the SM is trained, it is evaluated on both the test and training data, and compared to the original simulation output values. The normalized mean absolute error (MAE) is calculated and reported. To prevent overfitting, the PCE is run repeatedly with increased order to minimize the MAE until the difference between the test and training dataset are more than 5\%. In our case, this was at 4th order.\\
 A general workflow scheme for surrogate model creation from simulation data is depicted in \figref{fig:ML_schematic_inputbeam}.

To design and train neural networks we used the TensorFlow \cite{tensorflow2015-whitepaper}
machine learning framework and the hyperparameter optimization tools provided by Keras \cite{chollet2015keras}. These support automated tuning of the neural network hyperparameters, the used boundary values of which are given in \tabref{tab: hypers}. We underwent a new hyperparameter scan for each case, and  automatically selected the best hyperparameter configuration with minimized MAE for the training set. The obtained surrogate model can be saved and used for sensitivity studies and optimization.

\begin{table}
   \centering
   \caption{Hyperparameter boundaries for neural network tuning.}
   \vspace{1mm}
   \begin{tabular}{ll}
	Depth & 2 to 40\\
    Width & 3 to 160 \\
    Learning Rate & 0.1 to 0.0001 \\
    Batch Size & 8, 16, 32, 128 or 256\\
    Loss Function & Mean Square Error \\
    Epochs &  Up to 10000\\
    Activation Function & Relu\\
    L2 Regularization Penalty & 0.001 to 0.05\\
    Gaussian Noise & 0.01 to 0.1\\
   \end{tabular}
	
   \label{tab: hypers}
\end{table}

Based on the surrogate model, an optimization of the design variables with respect to the objectives using a generic optimizer algorithm can be performed, the result of which (SM output for the best found set of DVARs) can then be validated by the result of the corresponding PARMTEQM beam dynamics simulation.       

\section{Results}
\label{sec:results}

\subsection{Basic FODO Cell Example}

\begin{figure}[b!]
   \centering
    \includegraphics*[width=0.375\textwidth]{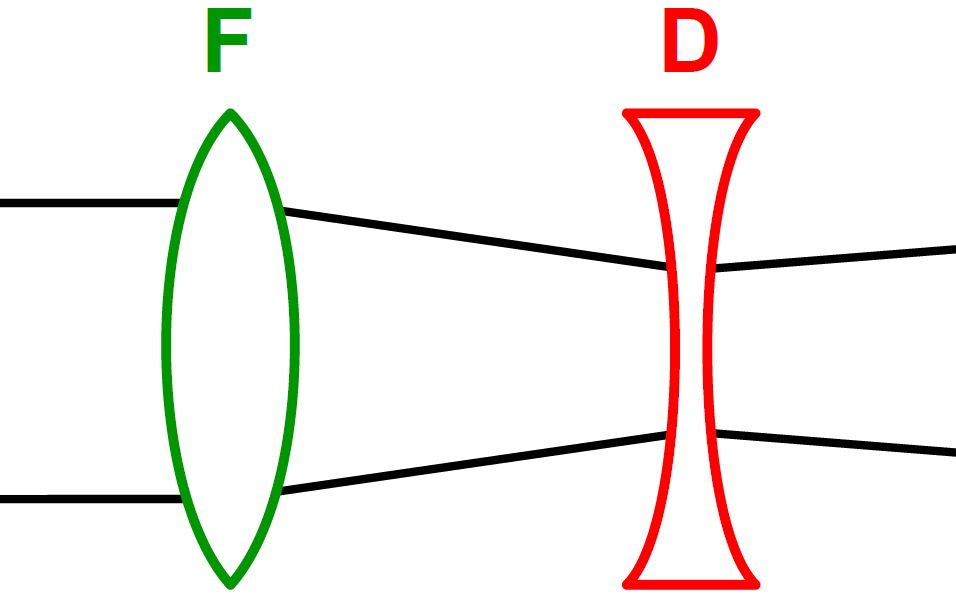}
    \caption{Schematic depiction of a FODO cell, showing a transverse projection of the beam envelope undergoing focusing (F), drift (O), defocusing (D) and drift (O).}
    \label{fig:FODO}
\end{figure}

The effects of a quadrupole magnet on an ion beam causes focusing on one transverse spatial axis, while leading to defocusing in perpendicular direction. However, using alternating quadrupoles in series can lead to a net focusing effect for the beam. In accelerator physics, one of the most basic examples of this is called a FODO cell, thus named for Focusing, drift (0), Defocusing, drift (0). This is seen in \figref{fig:FODO}.

In order to demonstrate the feasibility of using machine learning techniques to replicate accelerators, we started by reproducing the beam dynamics of this focusing/defocusing FODO lattice. This is the simplest and most basic example that still features similar transverse beam behavior as in RFQs but with greatly reduced overall complexity, and was therefore decided to be a good case to prove the proposed modeling concept.\\
We computed the FODO cell simulations in OPAL\cite{adelmann:opal}, using beam input parameters as summarized in \tabref{FODO_input_parameters}. As shown in \tabref{tab:SM_results}, the generated surrogate model of the FODO cell is capable of mapping the beam input parameters accurately to the values of the output emittances (both transversely and longitudinally) with MAEs of less than 1\%, regarding the test data set.

\begin{table}
   \centering
   \caption{Input beam design variables to the fixed FODO cell lattice generated using OPAL, and the range of their parameter space.}
   \vspace{1mm}
   \begin{tabular}{ll}
					corx & -0.5 to 0.5 \\
					cory & -0.5 to 0.5 \\
					Beam current [mA] & 2 to 10 \\
					RMS t [MeV deg] & 0.0001 to 0.0005 \\
					RMS x [m] & 0.001 to 0.005 \\
					RMS y [m] & 0.001 to 0.005 \\
   \end{tabular}
	
   \label{FODO_input_parameters}
\end{table}

\subsection{FODO Lattice with Varying Cell Parameters}
In addition to manipulating the beam input properties and simulating the beam through a fixed FODO cell, we also investigated the case of a variable hardware setup by using the focusing strengths K1 and K2 of the FODO cell quadrupole magnets as design variables. A summary of all design variables of the investigated system is given in \tabref{FODO_full_input_parameters}.\\

\begin{table}[h]
   \centering
   \caption{Design variables and range of their parameter space for the FODO lattice system with varying beam and cell parameters.}
   \vspace{1mm}
   \begin{tabular}{ll}
					Beam Current [mA] & -0.5 to 10 \\
							K1 [m$^{-2}$] & 4.2 to 4.8 \\
					K2 [m$^{-2}$] & 5.2 to 5.7 \\
					RMS t [MeV deg] & 0.0001 to 0.0005 \\
					RMS x [m] & 0.001 to 0.005 \\
					RMS y [m] & 0.001 to 0.005  \\

   \end{tabular}
	
   \label{FODO_full_input_parameters}
\end{table}

This scenario resulted in significantly larger errors compared to the fixed cell example where variation was restricted to the input beam properties. A more detailed discussion of this issue is given later in the discussion section of this paper. The yielded MAE values can again be found in \tabref{tab:SM_results}.

\subsection{Creating a Beam Dynamics Tuning Tool for an RFQ}

\begin{figure}
   \centering
    \includegraphics*[width=0.5\textwidth]{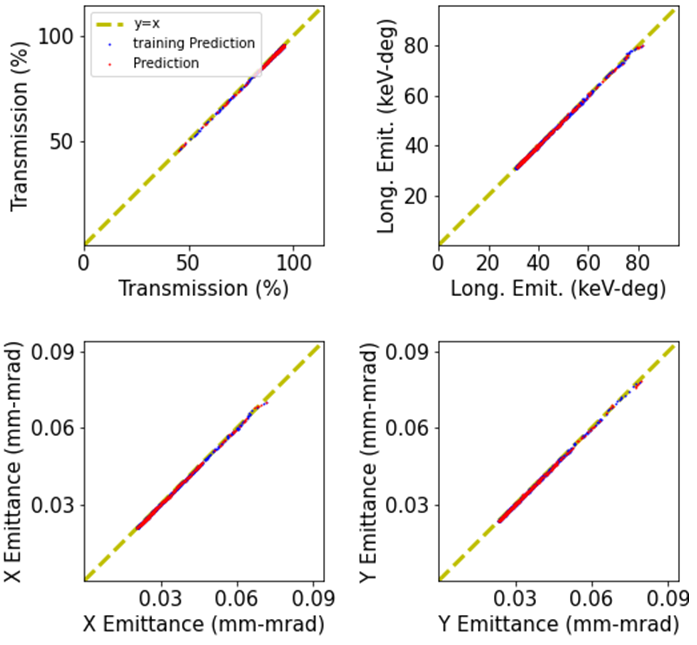}
    \caption{Neural network surrogate model predictions as function of the test data values for the variation of only the beam input Twiss parameters to a fixed RFQ (MAEs being well below 1\%).}
    \label{fig:twissmatch}
		\vspace{1mm}
\end{figure}

Next, we created a surrogate model with the aim to reproduce the beam dynamics behavior through the RFQ, given a fixed RFQ and variable LEBT input parameters.
 As summarized in \tabref{tab:SM_results}, a very high model accuracy could be achieved (using either PCE or NN) with values of the normalized MAEs typically being below 1\%, regarding transmission and emittances. Corresponding accuracy plots are shown in \figref{fig:twissmatch}.\\ 
 
 Because executing the surrogate models takes only about $7\cdot10^{-4}\,\text{s}$, given the used computer hardware and software specification, this method can be used to rapidly model the RFQ output for different inputs from the LEBT, allowing to compare simulations and commissioning data in real time. We have thus been able to create a real time, accurate tool to use during the commissioning phase of our RFQ.

 Furthermore, we were able to use the same surrogate model to optimize the input beam Twiss parameters ($\alpha$ and $\beta$) given a fixed RFQ setup.
 
Due to the high-fidelity of the achieved surrogate model, the intended optimization of the input beam Twiss parameters for RFQ injection could be performed using a Bayesian optimizer \cite{bayes}, with the SM as the test function and maximum output transmission and minimum output emittances as optimization objectives.

\begin{table}[h!]
   \centering
   \caption{Optimum set of Twiss parameters found by Bayesian optimizer based on the surrogate model output and corresponding predicted beam output parameters with comparison to PARMTEQM results.}
   \vspace{1mm}
   \begin{tabular}{lll}
					Input $\alpha$  & 2.55 & \\
					Input $\beta$ & 16.60 cm/rad &  \\
					\\
					 & SM output & PARMTEQM output \\
					Transmission [\%] & 95.5  & 95.3\\
					$\epsilon_s$ [MeV Deg] & 0.031 & 0.031\\
					$\epsilon_x$ [mm mrad] & 0.021 & 0.021\\
					$\epsilon_y$ [mm mrad] & 0.024 & 0.024\\
   \end{tabular}
	
   \label{tab:Twiss_optimization}
\end{table}

To cross check the optimization results based on the SM, the found optimum set of Twiss parameters was then used to validate the predicted SM output by PARMTEQM simulations. The optimum Twiss parameters found for a preliminary revised design of the IsoDAR RFQ are given in \tabref{tab:Twiss_optimization} together with the predicted beam output parameters by the SM and the corresponding PARMTEQM output. Deviations between the simulation and the SM prediction (= optimization result) are less than 0.2\,\% for both transmission and emittance values.

\subsection{Optimization of the entire RFQ Beam Dynamics Design on the Basis of Surrogate Models}
\begin{figure}[t!]
    \centering
    \includegraphics[width=1.0\columnwidth]{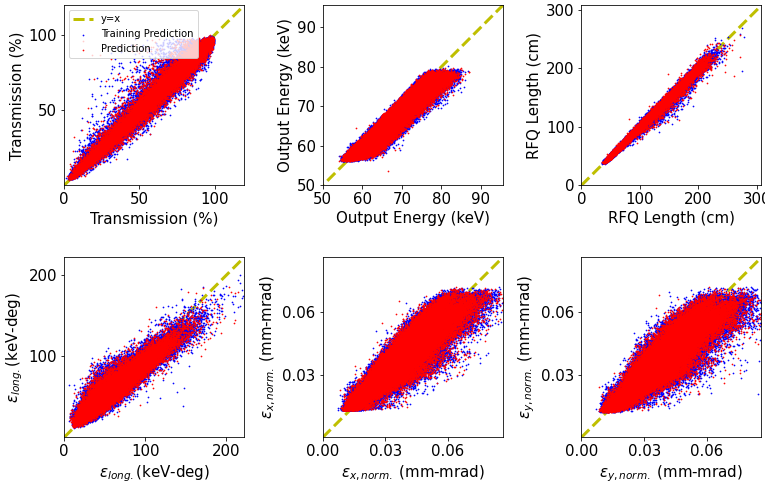}
    \caption{Surrogate model predictions as function of the test data values for full RFQ design variation by 14 DVARs and fixed beam input Twiss parameters.\label{fig:SM_accuracy}}
 
\end{figure}

Ultimately, we used the 14-DVAR RFQ model sample data set to train PCE and NN based models. Corresponding accuracy plots can be seen in \figref{fig:SM_accuracy} and achieved MAEs are again summarized in \tabref{tab:SM_results}.\\
Similar to the previous case, the obtained surrogate models execute much faster than their simulation counterparts. Whereas the calculation of a SM prediction takes around $10^{-3}\,\text{s}$, a corresponding physics beam dynamics simulation with PARMTEQM of a short IsoDAR type RFQ with an electrode length of around 1.3\,m consumes up to around 40\,s. With a sufficiently large design space, this reduces the time to find an optimized RFQ beam dynamics design, which otherwise typically would require a very high number of simulation iterations.

With MAEs of the predicted output emittances of up to 10\,\% (the MAEs for the transmission however being noticeably smaller) we found the Surrogate Models currently do not provide decent enough accuracy in any of the considered cases to perform a full RFQ design optimization. However, these computationally inexpensive surrogate models can be used to perform a rough pre-optimization with respect to the beam output objectives, providing a starting point for fine tuning optimizations using beam dynamics simulation tools. Using these methods combined reduces the total computational need of RFQ optimization and allows to quickly explore different possible qualitative solution approaches.

\begin{table*}[!hbt]
\centering
\caption{Comparison between mean average errors (MAEs) for surrogate models based on polynomial chaos expansion (PCE) and neural networks (NN) for different optimization cases and objectives.}
\includegraphics*[width=1.5\columnwidth]{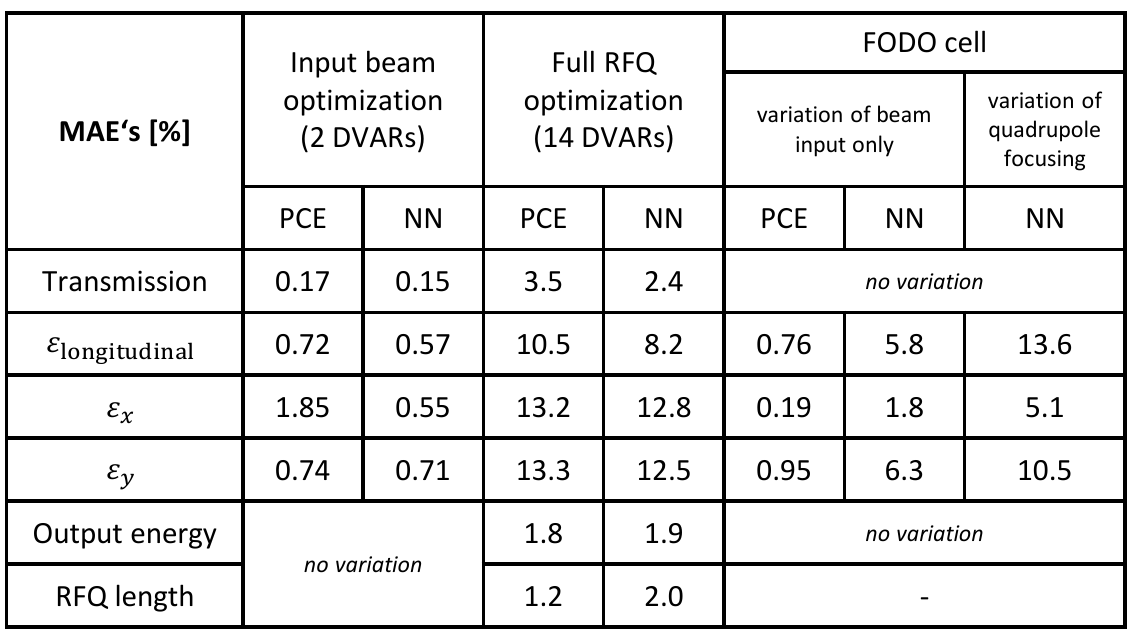}
\label{tab:SM_results}
\end{table*}

\section{Discussion}
\label{sec:discussion}

The created surrogate models quickly proved to be a reliable rapid-use tool for observing the effects of input beam variations on the output beam properties of a given RFQ. This has been a useful tool in optimizing the LEBT design, and could be as much as useful during commissioning and tuning of the LEBT/RFQ system.\\
However, the application of the developed techniques to the full RFQ beam dynamics design optimization proved problematic due to increased errors in predicted emittance whenever the space of design variables was expanded to include physical changes to the RFQ.\\

We found that highly accurate ($<$1$\%$ mean average error, MAE) surrogate models can be obtained for the optimization of only the input beam Twiss parameters (2 DVARs), as well as for a simplistic test case of modeling the beam dynamics in a FODO lattice under variation of the beam input parameters. For these simple cases, an optimization based on the surrogate model could be performed, with small deviations of the results to the beam dynamics simulation. In general, the use of neural networks (NN) leads to more accurate surrogate models compared to polynomial chaos expansion (PCE).\\

As shown in Fig.\,\ref{fig:SM_accuracy} and summarized in Tab.\,\ref{tab:SM_results}, models that include a structural changes of the accelerator system, such as variation of the FODO cell focusing strengths and the full RFQ optimization, suffer from much higher errors, especially regarding the emittances ($>$10$\%$). In none of the problematic cases did the error values improve significantly by switching off space charge (beam dynamics simulation with zero-current). When comparing the FODO cell example with the full RFQ optimization, it seems that the higher errors result not from a larger number of design variables, but are only introduced in case that the design variables affect the structure of the accelerator itself.\\

\begin{figure}[t!]
   \centering
   \includegraphics*[width=0.95\columnwidth]{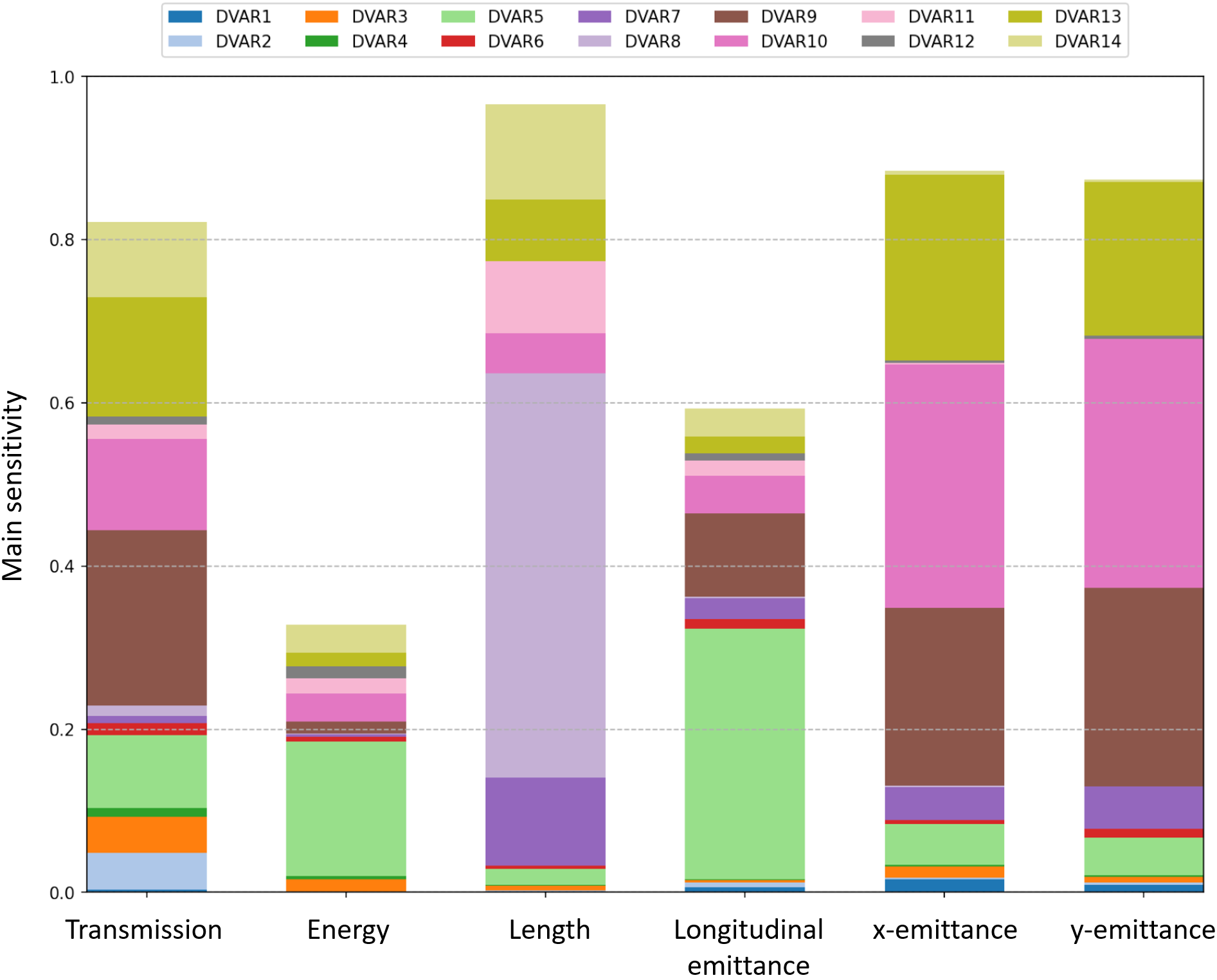}
   \caption{Sensitivity plot for the full RFQ optimization with 14 design variables (DVARs).} 
   \label{fig:SensitivityPlot}
 \end{figure}

As depicted in \figref{fig:SensitivityPlot}, the surrogate model lends itself to perform sensitivity analyses investigating the impact of DVAR variation on the optimization objectives.\\ 
Eventually, this allows for an evaluation of the cell properties parameterization model and to reduce the number of DVARs by omitting design variables with little effect on the crucial optimization objectives.\\
In case of our specific RFQ, the sensitivity chart (\figref{fig:SensitivityPlot}) reaveals that variation of DVARs 9, 10 and 13 (all relating to the function $\phi(z)$ for the synchronous phase) have the most significant influence on the transverse emittances, while the longitudinal emittance seems to be most sensitive to DVAR5 (value of the modulation factor $m(z)$ at the end of the exponential shaping section). Potential DVAR variations that might be omitted for the optimization procedure apparently relate to DVAR1 (value of the transverse focusing parameter $B(z)=\text{const.}$) and DVARs 2 and 4 (properties of $m(z)$ in the slow linear shaping section) as well as DVARs 3 and 6 (properties of $m(z)$ in the exponential shaping section).

What we have found is that surrogate models are able to closely reproduce simulations of beam dynamics in a fixed accelerator setup, however when the accelerator parameters change (e.g. the magnetic field strengths in the FODO cell or the configuration of the electric quadrupole field in the RFQ), the normalized mean average error in the emittances rises to ~12\%. While this error is too high to do a full optimization of the investigated system, surrogate modeling still proved useful to eliminate large areas of design space. With a reduced design space, the accelerator can be fine tuned using more accurate, computationally expensive codes in the region of interest.

Future work will investigate the use of other neural network structures that are not fully connected systems. It is possible that the errors may be further reduced by altering the structure of the neural network, while maintaining the high computational speed.

\section{Conclusion}
\label{sec:conclusion}
In this paper, we applied a recently developed 
surrogate modeling technique to the optimization of the beam output quality of RFQ linear accelerators for the first time.
We tested our method on a simple FODO cell (having similar transverse focusing properties)
first and on the IsoDAR RFQ thereafter.
To create the surrogate models, we used polynomial chaos expansion and deep neural networks.
We compared the results and found that we could very accurately predict the beam behaviour
from varying input beam parameters as it goes through a fixed accelerator structure, 
which initially was our main goal.
The trained model is intended to be used as an online feedback tool in the commissioning and tuning of 
the IsoDAR injector.
Furthermore, we found that, when we train the surrogate model on sets of hardware parameters
(i.e. many different design configurations of the investigated machine), we incur much higher training and validation errors.
We are in the process of investigating the cause of this effect, and we can already 
say that, in a comparison between beams with and without space charge, we do not see 
a difference.
Despite the large training errors (up to 10\,\%), the surrogate models 
trained on hardware design variables can be used to perform preliminary optimization of 
the design, reducing the model space, followed by a second iteration stage using 
high-fidelity physics simulations. Furthermore, Sobol's indices can be used to elucidate 
the influence of single design variables on the objectives, allowing to restrict design variations to the most crucial properties, thereby reducing the hyperspace of the input variables for the optimization process.

\begin{acknowledgments}
This work was supported by NSF grants PHY-1505858 and 
PHY-1626069, as well as funding from the Bose Foundation.
\end{acknowledgments}

\bibliographystyle{apsrev4-2} 
\bibliography{Bibliography}

\end{document}